\documentclass[aps,twocolumn,preprintnumbers,amsmath,amssymb,floats,citeautoscript]{revtex4-1}
\usepackage{graphicx}
\usepackage{bm}

\begin{document}
\preprint{Science {\bf 336}, 1554--1557 (2012).}
 
\title{A Sharp Peak of the Zero-Temperature Penetration Depth at Optimal Composition in BaFe$_2$(As$_{1-x}$P$_x$)$_2$}

\author{K. Hashimoto$^{1,*}$, K. Cho$^{2,3}$, T. Shibauchi$^{1,\dag}$, S. Kasahara$^{1,4}$, Y. Mizukami$^{1}$, R. Katsumata$^{1}$, Y. Tsuruhara$^{1}$, T. Terashima$^{4}$, H. Ikeda$^{1}$, M.\,A. Tanatar$^{2}$, H. Kitano$^{5}$, N. Salovich$^{6}$, R.\,W. Giannetta$^{6}$, P. Walmsley$^{7}$, A. Carrington$^{7}$, R. Prozorov$^{2,3}$, Y. Matsuda$^{1,\dag}$}

\affiliation{
$^1$Department of Physics, Kyoto University, Kyoto 606-8502, Japan}
\affiliation{
$^2$The Ames Laboratory, Ames, IA 5001, USA}
\affiliation{
$^3$Department of Physics \& Astronomy, Iowa State University, Ames, IA 50011, USA}
\affiliation{
$^4$Research Center for Low Temperature and Materials Sciences, Kyoto University, Kyoto 606-8501, Japan}
\affiliation{
$^5$Department of Physics and Mathematics, Aoyama Gakuin University, 5-10-1 Fuchinobe, Chuo-ku, Sagamihara, Kanagawa 252-5258, Japan}
\affiliation{
$^6$Loomis Laboratory of Physics, University of Illinois at Urbana-Champaign, 1110 West Green St., Urbana, IL 61801, USA}
\affiliation{
$^7$H. H. Wills Physics Laboratory, University of Bristol, Tyndall Avenue, Bristol, UK}
\affiliation{
$^*$Present address: Institute for Materials Research, Tohoku University, Sendai 980-8577, Japan}
\affiliation{
$^\dag$To whom correspondence should be addressed. E-mail: {\sf\small matsuda@scphys.kyoto-u.ac.jp} {\rm\small (Y.M.)}}
\affiliation{{\sf\small shibauchi@scphys.kyoto-u.ac.jp} {\rm\small (T.S.)}
}

\date{published June 22, 2012}

\begin{abstract}
{\bf
In a superconductor, the ratio of the carrier density, $n$, to their effective mass, $m^*$, is a fundamental property directly reflecting the length scale of the superfluid flow, the London penetration depth, $\lambda_L$. In two dimensional systems, this ratio $n/m^*$ ($\sim 1/\lambda_L^2$) determines the effective Fermi temperature, $T_F$. We report a sharp peak in the $x$-dependence of $\lambda_L$ at zero temperature in clean samples of BaFe$_2$(As$_{1-x}$P$_x$)$_2$ at the optimum composition $x = 0.30$, where the superconducting transition temperature $T_c$ reaches a maximum of 30\,K. This structure may arise from quantum fluctuations associated with a quantum critical point (QCP). The ratio of $T_c/T_F$ at $x = 0.30$ is enhanced, implying a possible crossover towards the Bose-Einstein condensate limit driven by quantum criticality.
}
\end{abstract}



\maketitle

\begin{figure}[t]
\includegraphics[width=\linewidth]{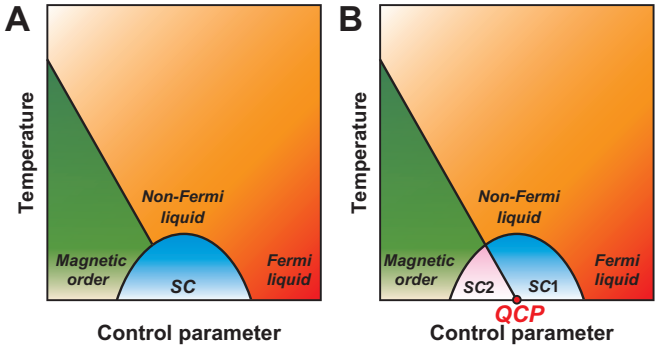}
\caption{Generic temperature vs. nonthermal control parameter phase diagram of iron-based superconductors illustrating two cases. ({\bf A}) Quantum criticality is avoided by the transition to the superconducting state. There is only one superconducting phase. ({\bf B}) A QCP lies beneath the superconducting dome. The QCP separates two distinct superconducting phases (SC1 and SC2). In the case of A, non-Fermi liquid behavior may appear above the dome if there is a QCP located along the axis of another control parameter that is independent of the control parameter shown on the abscissa. In the case of B, non-Fermi liquid behavior appears due to the QCP inside the dome.
}
\label{fig1}
\end{figure}

In two families of high temperature superconductors, cuprates and iron-pnictides, superconductivity emerges in close proximity to an antiferromagnetically ordered state, and the critical temperature $T_c$ has a dome shaped dependence on doping or pressure \cite{Stewart11,Hirschfeld11,Norman05}. What happens inside this superconducting dome is still a matter of debate \cite{Norman05,Dai09,Broun08}. In particular, elucidating whether a quantum critical point (QCP) is hidden inside it (Figs.\:\ref{fig1}A and B) may be key to understanding high-$T_c$ superconductivity \cite{Dai09,Broun08}. A QCP marks the position of a quantum phase transition (QPT), a zero temperature phase transition driven by quantum fluctuations \cite{Sachdev11}.

\begin{figure}[t]
\includegraphics[width=\linewidth]{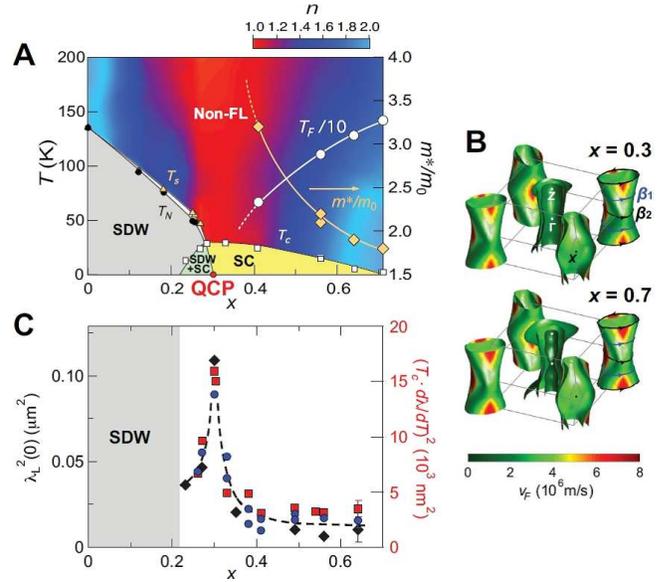}
\caption{({\bf A}) Phase diagram of BaFe$_2$(As$_{1-x}$P$_x$)$_2$. The transition to the SDW ground state at $T_N$ coincides or is preceded by the structural transition at $T_s$. With increasing $x$, $T_N$ decreases and goes to zero continuously at $x = 0.30$. The superconducting dome extends over a composition range $0.2 < x < 0.7$, with maximum $T_c = 30$\,K at $x = 0.30$. The red shaded region at around $x = 0.30$ represents the region where the exponent $n$ of the temperature dependence of the resistivity, $\rho_{dc}(T) = \rho(0) + aT^n$, is close to unity, which is a hallmark of a non-Fermi liquid (non-FL). The composition dependence of the effective Fermi temperature $T_F$ and renormalized mass $m^*/m_0$ determined by dHvA oscillations \cite{Shishido10} arising from the $\beta$ orbits (shown in B) are also plotted. ({\bf B}) Fermi surface of BaFe$_2$(As$_{1-x}$P$_x$)$_2$ with $x = 0.3$ and 0.7 from the band-structure calculation using density functional theory as implemented in the WIEN2K code \cite{Shishido10}. The Fermi surface consists of five quasi-cylindrical pockets, three hole pockets at the center of the Brillouin zone, and two electron pockets centered at its corners. The shading represents the in-plane Fermi velocity $v_F$. The flat parts of the outer electron sheets have high $v_F$ values. The lines represent the extremal $\beta$-orbits. ({\bf C}) Composition evolution of the square of the London penetration depth $\lambda_L^2(0)$ in the zero-temperature limit determined by three different methods: Aluminum coating method (black diamonds), microwave cavity perturbation technique (blue circles), and the low-temperature slope of the change of the penetration depth with temperature (red squares, right hand scale) shown in Fig.\:3. Different points for the same $x$ correspond to different crystals from the same batch. Error bars shown for $x=0.64$ represent typical experimental errors for all $x$, which involve the measurement errors \cite{SOM} and typical systematic uncertainties including the sample dependence. 
}
\label{fig2}
\end{figure}

The London penetration depth $\lambda_L$ is a property that may be measured at low temperature in the superconducting state to probe the electronic structure of the material, and look for signatures of a QCP. The absolute value of $\lambda_L$ in the zero-temperature limit immediately gives the superfluid density $\lambda_L^{-2}(0) = \mu_0e^2\sum_i n_i/m^*_i$, which is a direct probe of the superconducting state; here $m^*_i$ and $n_i$ are the effective mass and concentration of the superconducting carriers in band $i$, respectively \cite{SOM}. Measurements on high-quality crystals are necessary because impurities and inhomogeneity may otherwise wipe out the signatures of the QPT. Another advantage of this approach is that it does not require the application of a strong magnetic field, which may induce a different QCP or shift the zero-field QCP \cite{Moon09}.

BaFe$_2$(As$_{1-x}$P$_x$)$_2$ is a particularly suitable system for penetration depth measurements as, in contrast to most other Fe-based superconductors, very clean \cite{Shishido10} and homogeneous crystals of the whole composition series can be grown \cite{Kasahara10}. In this system, the isovalent substitution of P for As in the parent compound BaFe$_2$As$_2$ offers an elegant way to suppress magnetism and induce superconductivity \cite{Kasahara10}. Non-Fermi liquid properties are apparent in the normal state above the superconducting dome (Fig.\:\ref{fig2}A) \cite{Kasahara10,Nakai10} and de Haas-van Alphen (dHvA) oscillations \cite{Shishido10} have been observed over a wide $x$ range including the superconducting compositions, giving detailed information on the electronic structure. Because P and As are isoelectric, the system remains compensated for all values of $x$ (i.e., volumes of the electron and hole Fermi surfaces are equal).

As discussed in Ref.\:\onlinecite{Shishido10}, the normal-state electronic structure of BaFe$_2$(As$_{1-x}$P$_x$)$_2$ determined by dHvA experiments is significantly modified from that predicted by conventional density functional theory (DFT) band structure calculations. Figure \ref{fig2}A shows the composition evolution of the effective mass $m^*$ normalized by the free electron mass $m_0$ and the Fermi temperature $T_F = \epsilon_F/k_B = \frac{\hbar e^2}{2\pi k_B m^*} A_k$ for $x > 0.4$, determined from the dHvA oscillations corresponding to the extremal orbits on the outer electron Fermi surface ($\beta_1$ and $\beta_2$ orbits in Fig.\:\ref{fig2}B). Here $\epsilon_F$ is the Fermi energy and $A_k$ is the cross sectional area of the orbit. In contrast to the negligible $x$-dependence expected from the DFT calculations, a critical-like increase in $m^*$ accompanied by a strong reduction of $T_F$ is observed as the system is tuned towards the optimal composition from the overdoped side.

For a reliable determination of the absolute value of $\lambda_L(0)$ in small single crystals, we adopted three different methods \cite{SOM}. The first is the lower-$T_c$ superconducting film coating method \cite{Prozorov06,Gordon10,Prozorov11}, in which $\lambda_L(0)$ is determined from the frequency shift of a high precision tunnel diode oscillator \cite{Hashimoto10} (resonant frequency of $f\sim 13$\,MHz) containing the BaFe$_2$(As$_{1-x}$P$_x$)$_2$ crystal coated with an aluminum film ($T_c = 1.2$\,K) of known thickness and penetration depth.

The second is the microwave cavity perturbation technique, in which $\lambda_L(0)$ is determined from the measurements of surface impedance, $Z_s = R_s + iX_s$, by using a superconducting resonator ($f \sim 28$\,GHz) and a rutile cavity resonator ($f\sim 5$\,GHz), both of which have a very high quality factor $Q\sim 10^6$ \cite{SOM}. In all crystals, the residual surface resistance $R_s(0)$ at $T\rightarrow 0$\,K, which we determined by withdrawing the crystal from the rutile cavity at low temperature, is less than 0.3\% of $R_s$ just above $T_c$. This negligible residual $R_s(0)$ indicates almost perfect Meissner screening without any non-superconducting regions. In the superconducting state well below $T_c$, $\lambda_L(T)$ is obtained from the surface reactance via the relation $X_s(T) = \mu_0\omega\lambda_L(T)$. The absolute value of $X_s$ is determined from $Z_s$ and dc-resistivity $\rho_{dc}$ (measured separately by a conventional four contact technique) by the relation $R_s=X_s=\sqrt{\mu_0\omega\rho_{dc}/2}$ which holds in the normal state \cite{SOM}.

\begin{figure}[t]
\includegraphics[width=\linewidth]{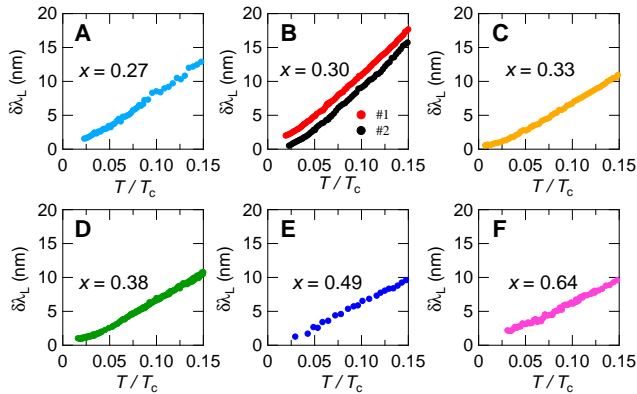}
\caption{Relative change of the penetration depth, $\delta\lambda_L(T) = \lambda_L(T)-\lambda_L(0)$, at low temperatures plotted against $T/T_c$ for different compositions from $x = 0.27$ ({\bf A}) to 0.64 ({\bf F}). For $x = 0.30$, data for two samples are shown, one of which (\#2) is shifted vertically for clarity.
}
\label{fig3}
\end{figure}

The third method uses the temperature dependent changes $\delta\lambda_L(T) = \lambda_L(T)-\lambda_L(0)$, measured by the tunnel diode oscillator down to $\sim 80$\,mK (Fig.\:\ref{fig3}). For all samples measured, covering a wide range of concentrations $0.26 \le x \le 0.64$, a quasi-$T$-linear variation of $\delta\lambda_L(T)$ is observed. This important result indicates that the presence of line nodes in the superconducting gap \cite{Hashimoto10} is a robust feature of this P-substituted system. This robustness is consistent with the nodes being on the electron sheets \cite{Yamashita11} rather than the hole sheets, as the electron sheets change relatively little with $x$ whereas the shape of the hole sheets changes substantially (Fig.\:\ref{fig2}B). A notable feature of the $T$-linear penetration depth is that the relative slope $\delta\lambda_L/d(T/T_c)$ is steepest for $x = 0.30$ (Fig.\:\ref{fig3}B). In general, this slope is determined by the Fermi velocity and the $\bm{k}$ dependence of the superconducting gap close to the node. Making the reasonable assumption that the gap structure evolves weakly across the phase diagram, the $x$ dependence of $\delta\lambda_L/d(T/T_c)$ will mirror that of $\lambda_L(0)$ \cite{SOM}.

Figure \ref{fig2}C shows the composition dependence of the squared in-plane London penetration length $\lambda_L^2(0)$ in the zero-temperature limit. Although different techniques can involve systematic errors \cite{Prozorov11}, all three methods give very similar $x$ dependencies. The most notable feature is the sharp peak in $\lambda_L^2(0)$ at $x = 0.30$, at about the same composition level where $T_c$ is maximal. The striking enhancement of $\lambda_L^2(0)$ is observed on approaching $x = 0.30$ from either side, and has been seen in multiple samples using different techniques. This reproducibility, combined with the above mentioned low $R_s(0)$, sharp superconducting transitions, and large heat capacity anomalies at all values of $x$ close to $x = 0.30$ \cite{SOM}, shows that the enhancement is not an experimental artifact associated with poor screening caused by non-bulk superconductivity. We attribute the peak in $\lambda_L^2(0)$ to the existence of a QCP at $x = 0.30$.

This result contrasts with the behavior found for the electron-doped iron-based superconductors, Ba(Fe$_{1-x}$Co$_x$)$_2$As$_2$, where a shallow minimum of $\lambda_L(0)$ at the optimum doping and a continuous increase on the underdoped side have been reported \cite{Gordon10,Prozorov11,Luan11}. This difference from the present case may be related to a greater degree of electronic disorder in the Fe layer caused by the Co doping \cite{Stewart11,Hirschfeld11}, which may smear out the singularity. The difference in the superconducting gap structure \cite{Hirschfeld11} as well as the additional of charge carriers from the Co doping, may also be a source of differences in the $x$-dependence of $\lambda_L(0)$. 

In cuprates, a QCP associated with the pseudogap formation has been suggested to occur at the hole concentration $p\sim 0.19$ inside the superconducting dome. We note, however, that there does not appear to be any evidence of mass divergence at this purported QCP and at this doping a broad minimum of $\lambda_L^2(0)$ was reported \cite{Tallon03}. An enhancement in $\lambda_L^2(0)$ has been observed at $p\sim 1/8$ \cite{Panagopoulos02}, but this is accompanied by a reduction of $T_c$, which is again different to the present case where the peak in $\lambda_L^2(0)$ coincides with the maximum $T_c$.

Our results may have general implications for the behavior of $\lambda_L^2(0)$ in strongly correlated superconducting systems. How strong electron correlations influence the condensed electron pairs in superconductors has been a long-standing issue \cite{Leggett65,Varma86,Gross86}. In fact, it has been pointed out that in an ordinary one-component Galilean invariant Fermi liquid, electron correlation effects do not cause the renormalization of $\lambda_L$ in the superconducting state \cite{Leggett65}. However, experimentally $\lambda_L^2(0)$ does appear to be enhanced in heavy-fermion superconductors, which contain interacting conduction electrons and local moments \cite{Varma86,Gross86}. The present results in BaFe$_2$(As$_{1-x}$P$_x$)$_2$ support this and suggest that in sufficiently clean systems electron correlation effects can lead to a striking renormalization of $\lambda_L^2(0)$.

We now discuss the consequences of a QPT inside the superconducting dome. Such a QPT implies that the non-Fermi liquid behavior indicated by the red region in Fig.\:\ref{fig2}A is most likely associated with a finite temperature quantum critical region linked to the QCP. Moreover, this transition immediately indicates two distinct superconducting ground states. In our system, the robust $T$-linear behavior of $\delta\lambda_L(T)$ on both sides of the purported QCP at $x = 0.30$ argues against a drastic change in the superconducting gap structure \cite{Hirschfeld11,Fernandes10}. The fact that the zero-temperature extrapolation of the antiferromagnetic transition $T_N(x)$ into the dome \cite{Nakai10} coincides with the location of the QCP (Fig.\:\ref{fig2}A) may indicate that the QCP separates a pure superconducting phase on the right and a superconducting phase coexisting with spin-density-wave (SDW) order on the left (Fig.\:\ref{fig1}B).

\begin{figure}[t]
\includegraphics[width=\linewidth]{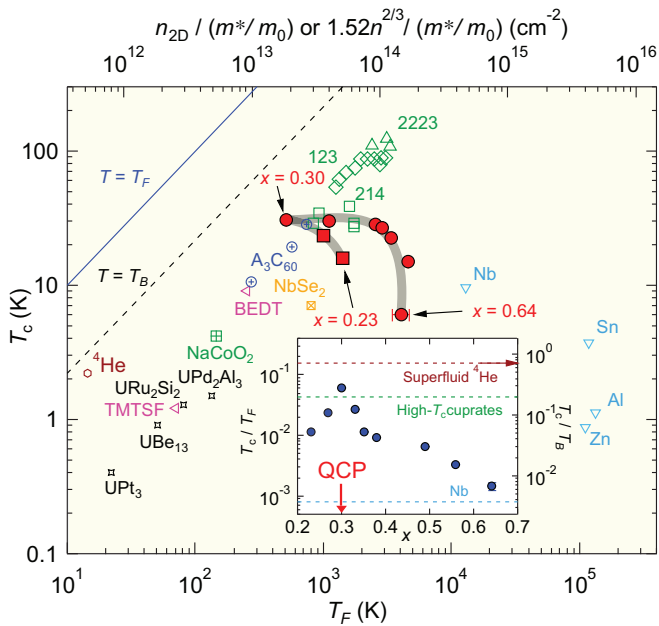}
\caption{Uemura-plot. $T_c$ is plotted as a function of effective Fermi temperature $T_F$ evaluated from $1/\lambda_L^2(0)$ for various superconductors ($n_{2D}/(m^*/m_0)$ for 2D and $1.52n/(m^*/m_0)$ for 3D systems) \cite{Uemura04}. We used an average of the Al-coating and microwave data for BaFe$_2$(As$_{1-x}$P$_x$)$_2$. The data for $x \ge 0.30$ (red circles) and for $x < 0.30$ (red squares) bridge a gap between the conventional superconductors such as Nb and cuprate high-$T_c$ superconductors such as (La,Sr)$_2$CuO$_4$ (214), YBa$_2$Cu$_3$O$_{7-\delta}$ (123), and Bi$_2$Sr$_2$Ca$_2$Cu$_3$O$_y$ (2223). $x = 0.30$ represents the data at the QCP. The dashed line is the BEC temperature for the ideal 3D boson gas. Inset: composition dependence of $T_c$ normalized by the Fermi temperature (left axis) or BEC temperature (right axis). Green and light blue dashed lines mark the $T_c/T_F$ values for underdoped cuprates 123 and for the conventional superconductor Nb. Brown arrow represents $T_c/T_B = 0.7$ for superfluid $^4$He.
}
\label{fig4}
\end{figure}

To place BaFe$_2$(As$_{1-x}$P$_x$)$_2$ in the context of other superconductors, Fig.\:\ref{fig4} plots $T_c$ as a function of the effective Fermi temperature $T_F$ for several types of compounds (Uemura plot); the red symbols correspond to various values of $x$ for BaFe$_2$(As$_{1-x}$P$_x$)$_2$ studied here, and the others are obtained from $\mu$SR measurements reported previously \cite{Uemura04}. Because the relevant Fermi surface sheets are nearly cylindrical, $T_F$ for 2D systems may be estimated directly from $\lambda_L(0)$ via the relation, $T_F=\frac{(\hbar^2\pi) n_{2D}}{k_B m^*}\approx \left(\frac{\hbar^2\pi}{\mu_0 e^2 d}\right) \lambda_L^{-2}(0)$, where $n_{2D}$ is the carrier concentration within the superconducting planes and $d$ is the interlayer spacing; $T_F=(\hbar^2/2)(3\pi^2)^{2/3} n^{2/3}/k_B m^*$ for three dimensional (3D) systems \cite{Uemura04}. The dashed line corresponds to the Bose-Einstein condensation (BEC) temperature for an ideal 3D boson gas, $T_B= \frac{\hbar^2}{2\pi m^* k_B}\left(\frac{n}{2.612}\right)^{2/3}\approx 0.0176 T_F$. In a quasi-2D system, this value of $T_B$ provides an estimate of the maximum condensate temperature. The evolution of $T_c$ with $T_F$ in the present system is in sharp contrast to that in cuprates, in which $T_c$ is roughly scaled by $T_F$. The inset of Fig.\:ref{fig4} depicts the $x$-composition dependence of Tc normalized by Fermi (or BEC) temperature, $T_c/T_F$ ($T_c/T_B$). In the large composition region ($x > 0.6$), $T_c/T_F$ is very small, comparable to that of the conventional superconductor Nb. As $x$ is decreased, $T_c/T_F$ increases rapidly, and then decreases in the SDW region after reaching the maximum at the QCP ($x = 0.30$). Notably, the magnitude of $T_c/T_B$ ($\approx 0.30$) at the QCP exceeds that of cuprates and reaches almost 40\% of the value of superfluid $^4$He.

The fact that $T_c/T_F$ becomes largest at the QCP indicates that the strongest pairing interaction is achieved at the QCP, implying that high-$T_c$ superconductivity is driven by the QCP. In a multiband system, we need to introduce the effective Fermi energy $\epsilon_F$ for each band, which is defined for electron (hole) bands as the energy of the highest occupied state relative to the bottom (top) of the band. Because the outer electron sheet with the highest Fermi velocity has the largest $\epsilon_F$ and hence the largest contribution to $\lambda_L^{-2}(0)$, the magnitude of $T_c/T_F$ in the other sheets are expected to be even larger. These results lead us to consider that in terms of $T_c/T_F$ the system is closer to the BCS-BEC crossover \cite{Chen05,Demelo93,Uemura09} than the cuprates.

\vspace{5mm}

\small
\noindent
{\bf Acknowledgments:} 
We thank R. Fernandes, S. Kivelson, H. Kontani, Q. Si, and Y. Yanase for valuable discussion. Supported by Grant-in-Aid for the Global COE program ``The Next Generation of Physics, Spun from Universality and Emergence'', Grant-in-Aid for Scientific Research on Innovative Areas ``Heavy Electrons'' from MEXT of Japan, KAKENHI from JSPS, and the EPSRC (UK). Work at the Ames Laboratory was supported by the Department of Energy-Basic Energy Sciences under Contract No. DE-AC02-07CH11358. Work at the University of Illinois was supported by the Center for Emergent Superconductivity, an Energy Frontier Research Center funded by the US Department of Energy, Office of Science, Office of Basic Energy Sciences under Award No. DE-AC0298CH1088.

\vspace{5mm}

\noindent
{\bf Supplementary Materials:} \\
www.sciencemag.org/cgi/content/full/336/6088/1554/DC1 \\
Materials and Methods\\
Supplementary Text\\
Figs. S1 to S5 \\
References [31-42]\\
30 January 2012; accepted 10 May 2012\\
10.1126/science.1219821

\clearpage

\renewcommand{\theequation}{S\arabic{equation}}
\setcounter{equation}{0}
\renewcommand{\thefigure}{S\arabic{figure}}
\setcounter{figure}{0}
\renewcommand{\thetable}{S\arabic{table}}
\setcounter{table}{0}
\makeatletter
\c@secnumdepth = 2
\makeatother


\begin{center}
{\large \bf Supplementary Materials}

\end{center}



\section{Sample characterization}

\begin{figure}[b]
\begin{center}
\includegraphics[width=\linewidth]{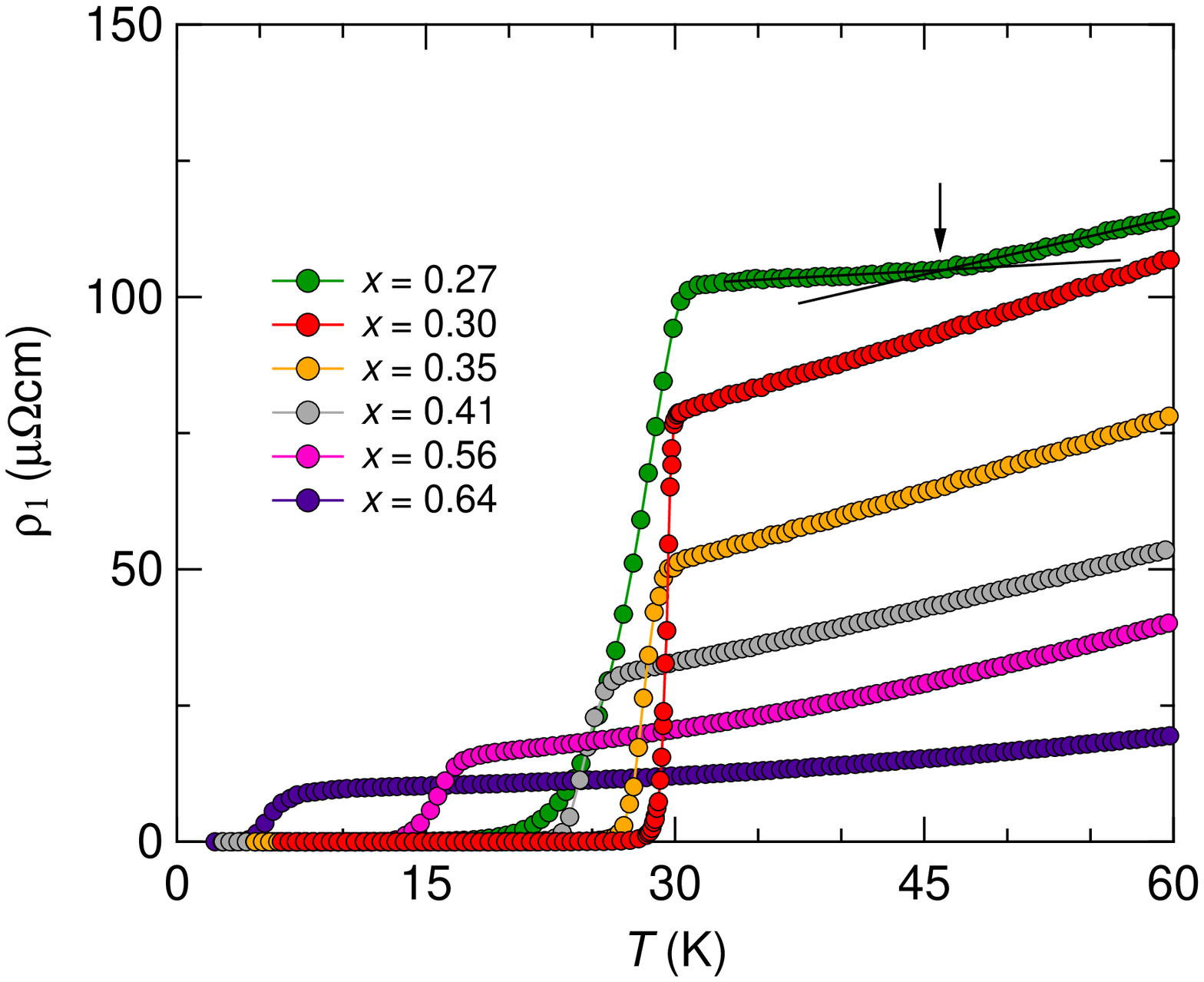}
\end{center}
\caption{Low-temperature microwave resistivity $\rho_1$ of BaFe$_2$(As$_{1-x}$P$_x$)$_2$ at 4.9\,GHz for various P-concentrations.  The arrow indicates a kink anomaly observed for $x=0.27$, which is attributed to the structural or SDW transition. The lines are guides for the eye.} \label{FigS1}
\end{figure}

\begin{figure}[t]
\begin{center}
\includegraphics[width=\linewidth]{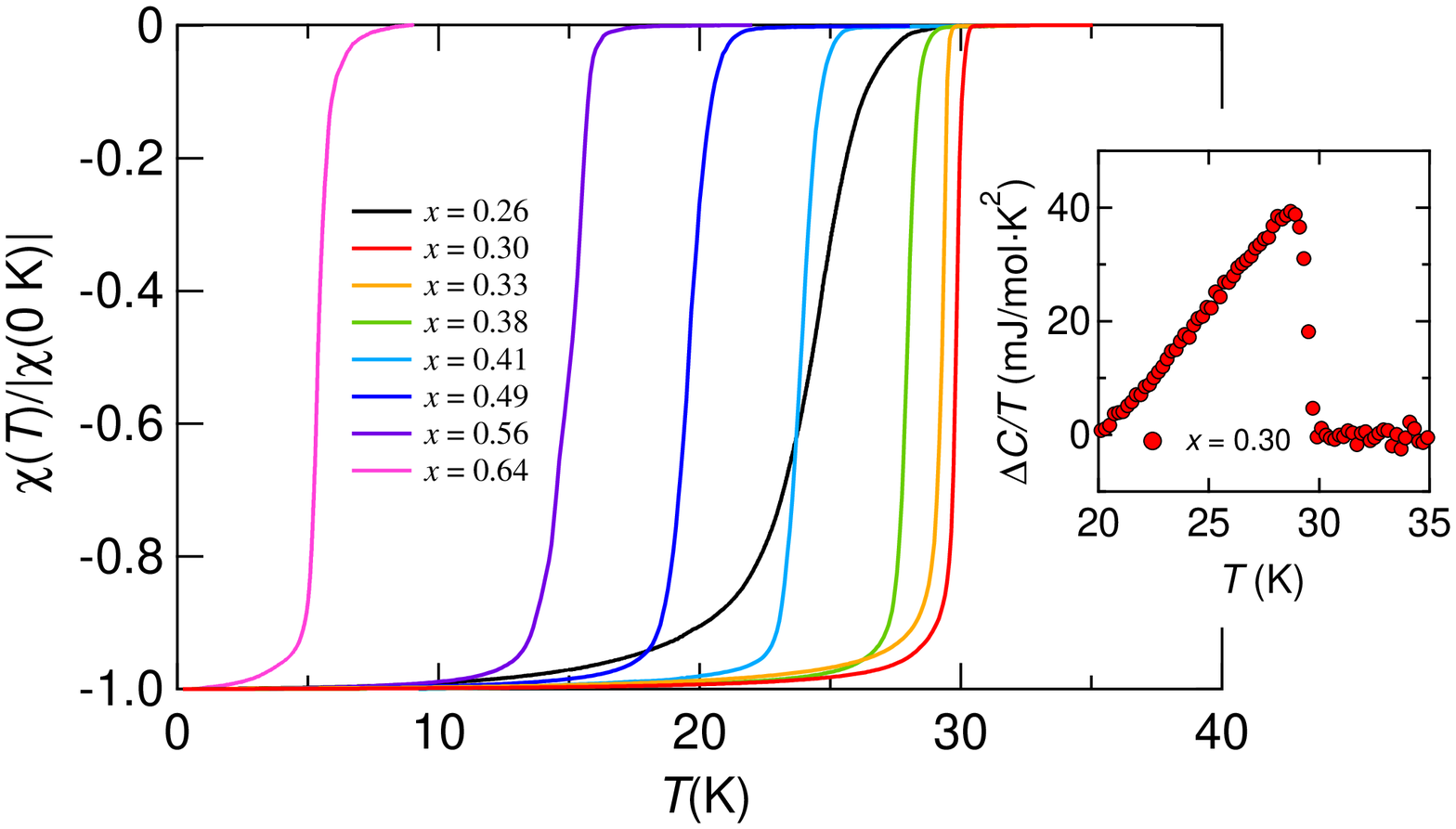}
\end{center}
\caption{Normalized ac susceptibility of BaFe$_2$(As$_{1-x}$P$_x$)$_2$ with different P concentrations $x$ ($0.26 \leq x \leq 0.64$) measured from the frequency shift of the tunnel diode oscillator. The inset shows the specific heat $C$ with the normal state contribution subtracted for the same $x = 0.30$ sample as was used for the temperature dependent penetration depth measurements (Fig.\:3).  The normal state contribution was estimated by extrapolating $C/T$ measured above $T_c$.  The measurements were made using a modulated temperature method [32], 
which has high resolution but poor absolute accuracy for small samples.  The absolute value of $\Delta C$ was therefore estimated by normalising the total specific heat to that measured on a similarly doped sample by a dc technique [18]. 
The uncertainty in $\Delta C$ is $\sim \pm 20$\% which is determined mostly by uncertainties in the addenda contribution.}
\label{FigS2}
\end{figure}

High-quality single crystals of BaFe$_2$(As$_{1-x}$P$_x$)$_2$ were grown by the self-flux method [11]. 
For the crystal structure analysis, X-ray diffractometry was carried out using four-circle diffractometer (MXC$^\chi$, MacScience). We find that the composition dependence of  the lattice parameters $a$, $c$, and the pnictogen height $h_{\rm Pn}$ obey Vegard's law, which allows us to determine the P-concentration $x$ within the error of $\pm 0.006$. In this study, BaFe$_2$(As$_{1-x}$P$_x$)$_2$ single crystals in a wide range of P-concentrations ($0.23 \leq x \leq 0.64$) were grown for the measurements of the London penetration depth. Typical size of the crystals is $\sim 200\times200\times10~\mu$m$^3$.  The crystals were cleaved just before the measurements to obtain the clean and flat surfaces.  Figure\:\ref{FigS1} shows the temperature dependence of  the microwave resistivity $\rho_1=2R_s^2/\mu_0\omega$ at 4.9\,GHz for the crystals used in this study, which coincides with the dc resistivity $\rho_{dc}$ in the normal state.  For $x = 0.27$, a kink anomaly associated with the structural or SDW transition is observed at 47\,K, which is consistent with the previous studies [11]. 
For $x=0.30$ at the optimal P-concentration, $\rho_1$ shows a $T$-linear  dependence, which is the hallmark of non-Fermi liquid behavior.  For $x=0.64$, $\rho_1$ shows a $T^2$-dependence, indicating a recovery of the Fermi-liquid behavior in the high concentration regime.  All these $\rho_1(T)$ behaviors are consistent with the previous $\rho_{dc}$ measurements [11]. 
Figure\:\ref{FigS2} shows the normalized ac susceptibility measured by the tunnel diode oscillator.  A very sharp superconducting transition is observed for $x\geq0.30$, demonstrating the high quality and homogeneity. The almost perfect superconducting volume fraction in the sample of $x = 0.30$ is supported by the large heat capacity anomaly (inset of Fig.\:\ref{FigS2}). The superconducting transition becomes broader for $x<0.30$, because the steep slope $dT_c/dx$ in the phase diagram (Fig.\:2A) implies that $T_c$ is very sensitive to small variations of $x$ in the low concentration regime. The bulk $T_c$ is defined by the temperature at which superfluid density becomes zero extrapolated from low temperatures.  We note that this $T_c$ coincides with the temperature at which specific heat and thermal expansion coefficients jump [31] 
and dc resistivity vanishes.

\section{Experimental techniques}

\subsection{Tunnel diode oscillator technique}

The tunnel diode oscillator (TDO) technique is able to extract the relative change of the penetration depth with temperature, $\delta\lambda(T) \equiv \lambda(T) - \lambda(0)$. The penetration depth measurements down to very low temperatures have been performed by using a high resolution radio frequency susceptometer based on a self-resonant tunnel diode circuit operating at $13$\,MHz [33], 
which is mounted on a $^3$He-$^4$He dilution refrigerator (base temperature $\sim 80$\,mK) or  $^3$He refrigerator (base temperature $\sim 300$\,mK). 

The TDO operates with an extremely small $ac$ probe field ($H_{ac} < 10$\,mOe) so that the sample is always in the Meissner state.  To measure the in-plane penetration depth, $H_{ac}$ is applied along the crystal $c$ axis, which generates the supercurrents in the plane.  The change in the resonant frequency $\delta f$ is proportional to the change in the penetration depth $\delta \lambda$, $\delta f=G \delta \lambda$. The calibration factor $G$ is determined from the geometry of the sample, and the total perturbation to the resonant frequency due to the sample, found by withdrawing the sample from the coil at low temperature [34]. 
The sample is mounted on a sapphire rod, the other end of which is glued to a copper block on which a RuO$_2$ or Cernox thermometer is mounted. The sample and rod are placed inside a solenoid which forms part of the resonant tank circuit.

\subsection{Microwave cavity perturbation technique}

\begin{figure}[b]
\begin{center}
\includegraphics[width=\linewidth]{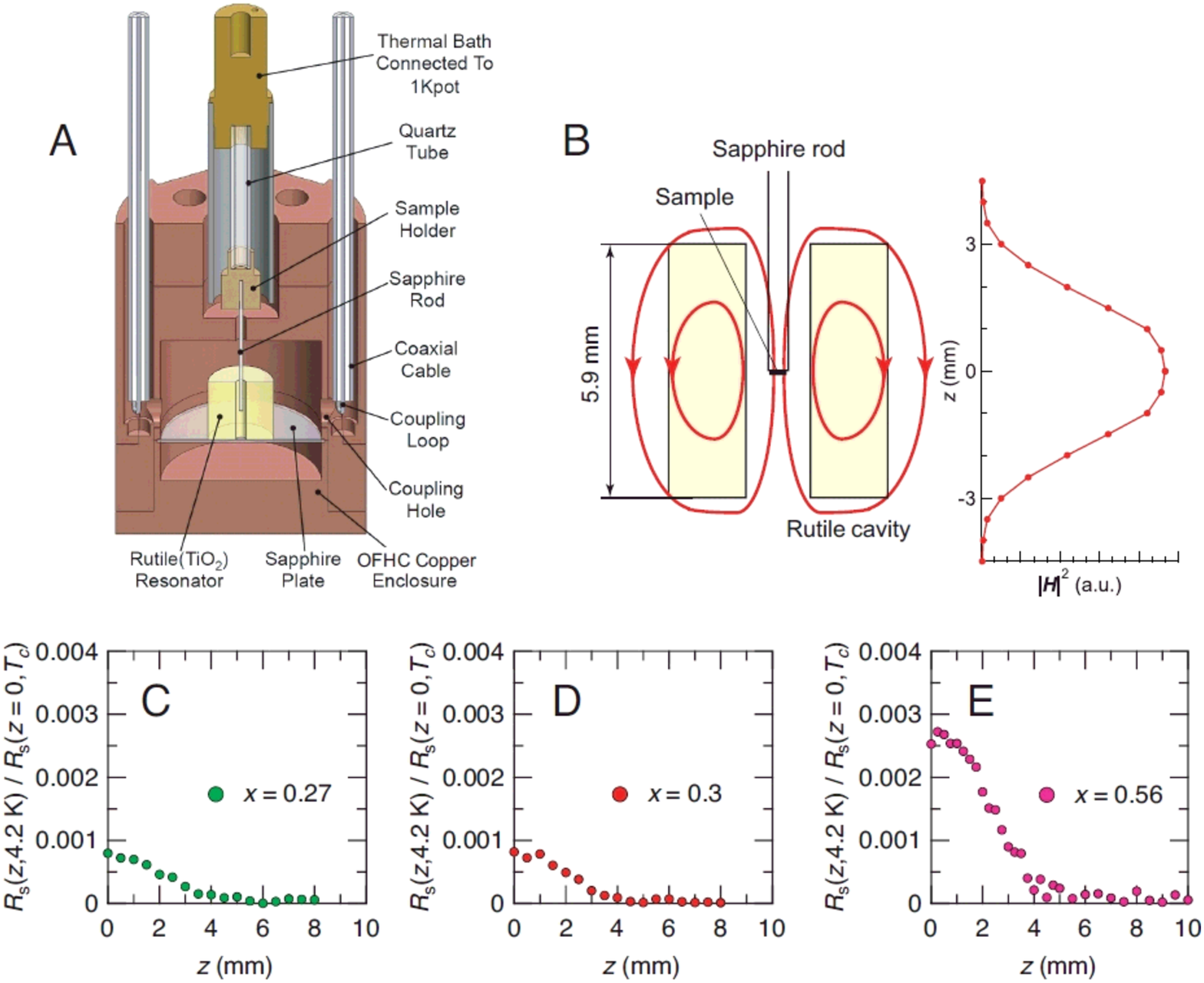}
\end{center}
\caption{({\bf A}) Schematic of the rutile cavity resonator. The rutile resonator is placed inside the OFHC copper enclosure. ({\bf B}) Left panel:  an illustration of the magnetic-field distribution of the TE$_{011}$ mode. Right panel: $z$ dependence of $|H|^2$. ({\bf C})--({\bf E}) The residual surface resistance at 4.2\,K normalized by the value just above $T_c$ at $z=0$ is plotted as a function of $z$ for the samples of $x=0.27$ (C), $x=0.30$ (D), and $x=0.56$ (E).
}
\label{FigS3}
\end{figure}

The penetration depth can also be determined by the measurements of the microwave surface impedance, $Z_s=R_s+iX_s$, where $R_s$ is the surface resistance and $X_s$ is the surface reactance.  
We have measured $Z_s$  by high-resolution microwave surface-impedance probes that use cavity perturbation of superconducting and dielectric resonators.  A cylindrical superconducting cavity resonator  with operating frequency of 27.8\,GHz  in the TE$_{011}$ mode is made of Pb-plated oxygen free high conductivity (OFHC) copper [35,36]. 
A cylindrical dielectric cavity resonator with operating frequency of 4.9\,GHz in the TE$_{011}$ mode is constructed from a high-permittivity single crystal of rutile (TiO$_2$) with high dielectric constant  ($\epsilon \approx 120$) [37]. 
Both resonators  have quality factors  in excess of $10^6$.

The superconducting cavity resonator is immersed in superfluid $^4$He at 1.6\,K and the temperature of the whole resonator is stabilized within $\pm0.5$\,mK.   Figure\:\ref{FigS3}A shows the schematic of the dielectric  cavity resonator. The magnetic flux is mainly confined in the rutile, which reduces the conduction current losses in the walls of the resonator enclosure.  Figure\:\ref{FigS3}B illustrates the magnetic field structure of the TE$_{011}$ mode.  The temperature of the rutile is stabilized within $\pm0.5$\,mK.  The crystal was mounted on a sapphire hot finger and placed at the antinode of the microwave magnetic field $\bm{H}_\omega$ ($\parallel c$ axis) so that the shielding currents $\bm{I}_\omega$ are excited in the $ab$ plane (Fig.\:\ref{FigS3}B).  The inverse of quality factor $1/Q$ and the shift in the resonance frequency are proportional to the surface resistance $R_s$ and the change in the surface reactance $\delta X_s=X_s(T)-X_s(0)$, respectively.  The sample stage is movable along the vertical axis $z$ of the sapphire rod from the center of the cavity ($z=0$).  By withdrawing the sample completely from the resonator, the background microwave absorption of the resonator can be measured directly, allowing the absolute microwave absorption of the sample to be determined {\it in situ}.   Figures\:\ref{FigS3}C--E show the $z$-dependence of the residual surface resistance $R_s$ at 4.2\,K normalized by the value just above $T_c$ at $z=0$ for three samples with different compositions.   The residual surface resistance at $4.2$\,K in all crystals is less than 0.3\% of $R_s$ at $T_c$, demonstrating nearly perfect Meissner state with negligibly small non-superconducting region.  We note that even in the low-concentration samples with broader transitions the residual $R_s$ value is comparable to that in optimal composition (Fig.\:\ref{FigS3}C).  This ensures that $\lambda_L(0)$ values in these samples give correct estimate of intrinsic superfluid density.

\section{Determination of the absolute value of the zero temperature London penetration depth}

\subsection{Al-coating method}

\begin{figure}[t]
\begin{center}
\includegraphics[width=\linewidth]{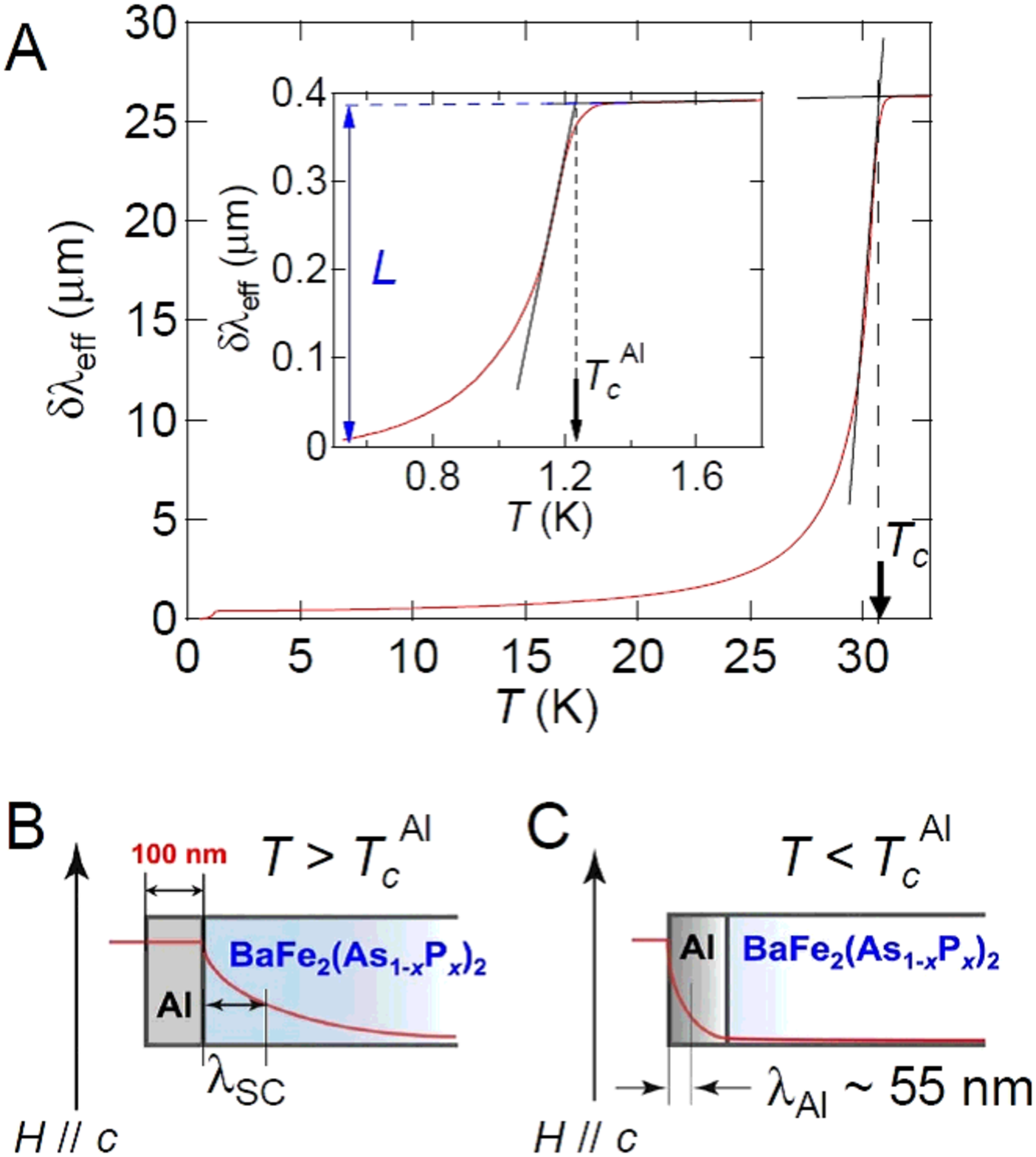}
\end{center}
\caption{  ({\bf A}) The temperature dependence of the effective penetration depth for the optimally doped BaFe$_2$(As$_{0.7}$P$_{0.3}$)$_2$ sample coated with Al film. The inset shows the low-temperature behavior.  The overall variation of the penetration depth below the Al transition, denoted as $L$, is used for the calculation of $\lambda(0)$. ({\bf B} and {\bf C}) Schematic of the field penetration in Al-coated sample at $T>T_c^{\rm{Al}}$ (B) and $T<T_c^{\rm{Al}}$ (C). }
\label{FigS4}
\end{figure}

To determine the absolute value of the penetration depth,  we used  the lower-$T_c$ superconducting film coating method [38]. 
We determined $\lambda(0)$  from the the frequency shift of the TDO containing the BaFe$_2$(As$_{1-x}$P$_x$)$_2$ crystal whose surfaces are coated with a thin film of aluminum (Al) having a lower critical temperature ($T_c^{\rm{Al}}= 1.2$\,K) and a known value of the penetration depth $\lambda_{\rm{Al}}(0) = 55$\,nm [13,14]. 
Figure\:\ref{FigS4}A shows the temperature dependence of the penetration depth for $x=0.3$ sample coated with the Al film of thickness of $\sim 100$\,nm.  Above $T_c^{\rm{Al}}$ (Fig.\:\ref{FigS4}B), the normal-state skin depth of Al ($\delta_{\rm{Al}}\sim 75\,\mu$m for $\rho_0^{\rm{Al}} = 10\,\mu\Omega$cm at 13\,MHz) is much larger than the thickness of the Al film.  As a result, the effective penetration depth into both the Al film and coated superconductor, $\lambda_{\rm{eff}}(T)$, is almost identical to the penetration depth $\lambda(T)$ before coating Al on the sample. On the other hand, when the Al film is superconducting (Fig.\:\ref{FigS4}C), Al acts together with the coated superconductor to screen the magnetic fields. Then, the effective magnetic penetration depth $\lambda_{\rm{eff}}(T)$ for $T < T_c^{\rm{Al}}$ can be given by
\begin{equation*}
\lambda_{\rm{eff}}(T) = \lambda_{\rm{Al}}(T)\frac{\lambda(T)+\lambda_{\rm{Al}}(T)\tanh{\frac{t}{\lambda_{\rm{Al}}(T)}}}{\lambda_{\rm{Al}}(T)+\lambda(T)\tanh{\frac{t}{\lambda_{\rm{Al}}(T)}}},
\label{Al_coated_eq}
\tag{S1}
\end{equation*}
where $t$ is the thickness of the Al film and $\lambda(T)$ is the penetration depth of the coated superconductor and $\lambda_{\rm{Al}}(T)$ is the penetration depth of the Al film [14]. 
Thus we can extract the absolute penetration depth of the coated superconductor. The overall penetration depth change below the Al transition, $L\equiv\delta\lambda_{\rm{eff}}(T)=\lambda_{\rm{eff}}(T)-\lambda_{\rm{eff}}(T_{\rm{min}})$, is used for the calculation of $\lambda(0)$, which gives $\lambda(0)=330$\,nm for $x=0.3$.   The typical error bar for $\lambda_L(0)$ measurements by this technique is $\pm$15\%.  The obtained results for the other concentrations $x$ are plotted in Fig.\:2C (black diamonds).

\begin{figure}[t]
\begin{center}
\includegraphics[width=\linewidth]{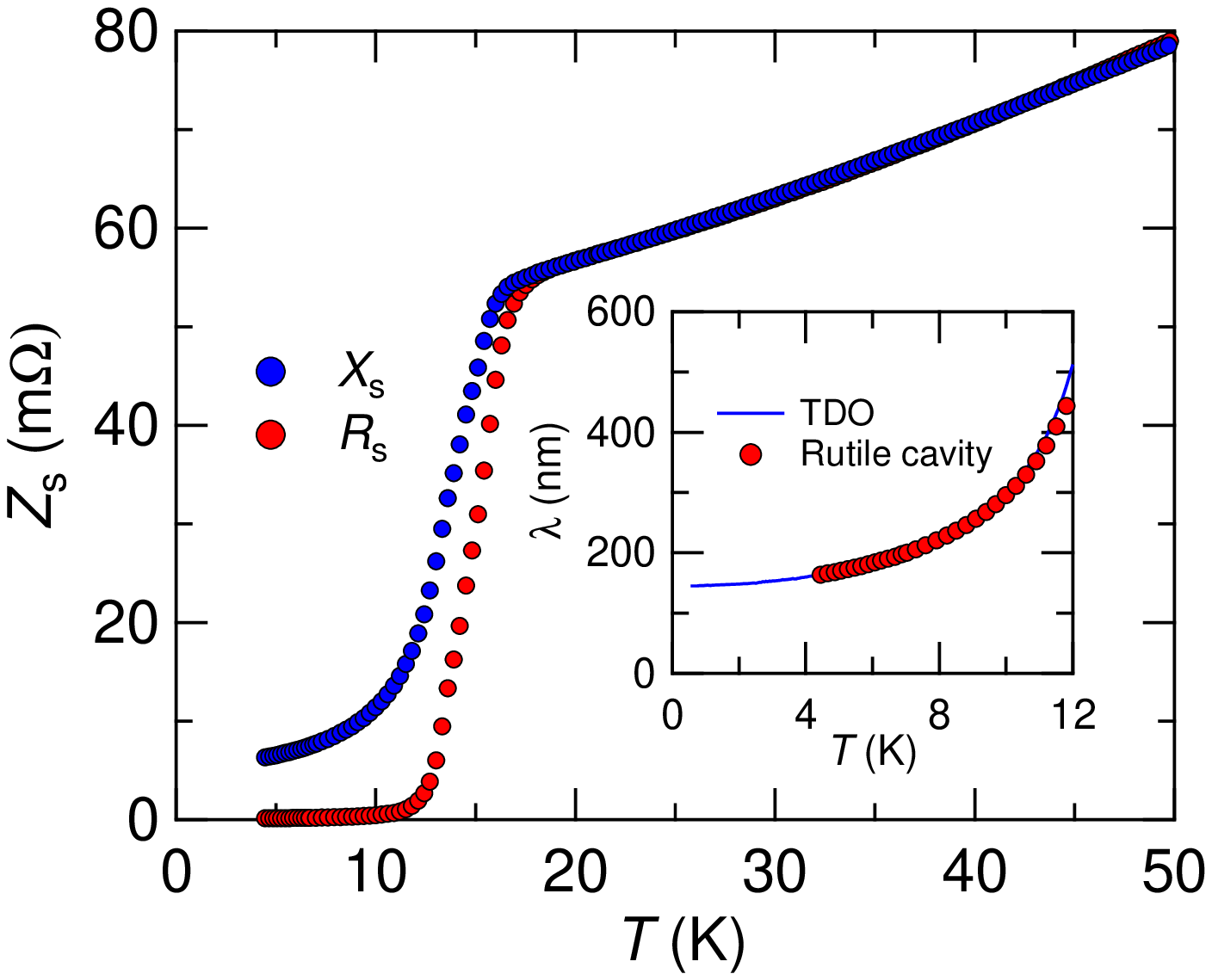}
\end{center}
\caption{Surface resistance $R_s$ and reactance $X_s$ as a function of the temperature at 4.9\,GHz for $x=0.56$. In the normal state, $R_s=X_s$ is observed. Inset shows the temperature dependence of the absolute penetration depth. Solid line represents the TDO data measured down to 80\,mK on the same single crystal.}
\label{FigS5}
\end{figure}

\subsection{Surface impedance measurements}

Figure\:\ref{FigS5} shows the temperature dependence of the surface resistance $R_s$ and reactance $X_s$ for the $x=0.56$ sample at 4.9\,GHz.  The low residual $R_s$ value in the low-temperature limit again demonstrates the high quality of the crystal.  The transition in microwave $R_s$  is broader than that in dc resistivity data since the applied GHz microwave  excites additional quasiparticles particularly just below $T_c$.  In the superconducting state well below $T_c$, $\lambda(T)$ is obtained  via the relation $X_s(T)=\mu_0\omega \lambda(T)$.    In the normal state, at the frequencies $\omega\tau \ll 1$, the surface resistance is equal to the surface reactance in the so-called Hagen-Rubens limit, $R_s=X_s=\sqrt{\mu_0\omega\rho_{dc}/2}$.   As shown in Fig.\:\ref{FigS5}, the relation $R_s=X_s$ holds in the normal state,  indicating the Hagen-Rubens limit in the present frequency region.   The absolute value of $X_s$ is thus determined by the dc-resistivity $\rho_{dc}$ in the normal state. The inset of Fig.\:\ref{FigS5} shows the temperature dependence of the absolute value of the penetration depth for the $x=0.56$ sample, demonstrating a good agreement with the TDO data shifted by a constant value. $\lambda(T)$ was extrapolated down to zero temperature by using the TDO data measured down to 80\,mK, which gives $\lambda(0)=135$\,nm.   The typical error bar for $\lambda_L(0)$ measurements by this technique is $\pm$15\%. The obtained results are summarized in Fig.\:2C (blue circles).

\subsection{Temperature dependent changes $\delta \lambda_L(T)$ }

The quasi-$T$-linear variation of $\delta \lambda_L(T)=\lambda_L(T)-\lambda_L(0)$ is observed in a wide range of $x$, indicating line nodes in the gap.  This temperature linear slope is related to $\lambda_L(0)$ and the superconducting energy gap magnitude $\Delta$.  For a $d$-wave nodal superconductor it is found that [13] 
\begin{equation*}
\frac{\delta \lambda_L(T)}{\lambda_L(0)}=\frac{\ln 2}{\Delta}k_BT.
\tag{S2}
\end{equation*}
If we assume this formula and a constant ratio $\Delta/k_BT_c$ (which in the simplest case corresponds to 2.14 for $d$-wave and 1.76 for $s$-wave), then $\lambda_L^2(0)$ can be estimated from the measured $d\lambda/dT$ and $T_c$. 
For example, when we use the $s$-wave ratio for $x=0.3$, $d\lambda/d(T/T_c)=125\pm10$\,nm, giving $\lambda_L=320\pm30$\,nm. This compares to 330\,nm obtained from the Al coating method and 300\,nm from the surface impedance methods. 
The red squares in Fig.\:2C are plotted in a scaled right axis, such that $\lambda_L^2(0)$ estimated by these simple assumptions (with $\Delta/k_BT_c=1.76$) from $(T_c\cdot d\lambda/dT)^2$ can be compared with the values in the left axis obtained directly by the other methods. Although the relation between $d\lambda/d(T/T_c)$ and $\lambda_L(0)$ in general depends on the number of nodes, the rate at which the gap opens near the node ($d\Delta/dk$) and also the relation between $\Delta$ and $T_c$, the correspondence with this simple estimate is striking. In any case, as long as these factors do not change sharply at the QCP it the observed strong increase in $d\lambda/d(T/T_c)$ close to the QCP is strong evidence supporting the other measurements which indicate a sharp peak in  $\lambda_L(0)$  at this composition.

\section{Microscopic coexistence of superconductivity and SDW order in the underdoped regime}

There has been much debate as to whether superconductivity coexists microscopically with other ordering not only in pnictides but also in cuprates, heavy fermions and other exotic superconductors, because of the difficulty to distinguish this from mesoscopic phase separation. The present results that the QCP separates two distinct superconducting phases strongly suggest that in the underdoped region, superconductivity and SDW coexist on a microscopic level, but compete for the same electrons in the underdoped region. This competition is evidenced by the overall larger $\lambda_L(0)$ values in the SDW side of the QCP than the other side (Fig.\:2C), corresponding to the smaller volume of Fermi surface due to partial SDW gapping.  The microscopic coexistence is also supported by the enhancement of $\lambda_L^2(0)$ on approaching the QCP from the SDW side, which is not expected in the case of phase separation [14,39]. 

\section{Microscopic interpretation of the magnetic penetration depth}
The magnetic penetration depth $\lambda$ is defined via the penetration of field into the surface of a superconductor
\[
\lambda = \frac{1}{B_0} \int_0^\infty B d\bm{r}.
\]
In a clean local, BCS superconductor, at $T=0$, the $x$ component of $\lambda$ can be evaluated by the following integral (in SI units) of the Fermi velocity $\bm{v}$ over the \textit{complete} Fermi surface $\bm{S}$ [40]
\begin{equation*}\label{Eq:lam}
    \lambda_x^{-2}(0)=\frac{\mu_0e^2}{4\pi^3\hbar}\int \frac{v_x^2}{|v|} d\bm{S}.
\tag{S3}
\end{equation*}
Here $v_x$ is the $x$ component of the Fermi velocity.  This equation is closely related to that for the normal state conductivity, $\bm{\sigma}$
\begin{equation*}
    \sigma_x=\frac{e^2}{4\pi^3\hbar}\int \frac{v_x^2 \tau}{|v|} d\bm{S}
\tag{S4}
\end{equation*}
where $\tau$ is the scattering time which in general depends on $\bm{k}$.  In the free electron, isotropic scattering approximation, these two equations can be reduced to
$ \lambda^{-2}(0) = \mu_0 n e^2/m$ and $\sigma = ne^2\tau/m$, where $m$ is the mass of the electron and $n$ is the density of electrons, and so it is common in the literature to abbreviate these integrals to the quantity $n/m^*$, where
\begin{equation*}
\frac{n}{m^*} = \frac{1}{4\pi^3\hbar} \int \frac{v_x^2}{|v|} d\bm{S}. \nonumber
\end{equation*}
For a multiband material, a separate $n/m^*$ can be evaluated for each sheet of Fermi surface and the contributions added to give the total $\lambda_x^{-2}(0)$.  The `mass' $m^*$ here is closely related to the cyclotron mass $m_c^*$ which is measured in de Haas-van Alphen effect experiments
\begin{equation*}
m_c^* = \frac{\hbar^2}{2\pi} \frac{dA}{d\varepsilon}, \nonumber
\end{equation*}\\
where $A$ is the Fermi surface (extremal) cross-sectional area and $\varepsilon$ is the quasiparticle energy.  Both masses are renormalized by the same factor when the Fermi velocity is renormalized by many-body effects.  This assumes that `back-flow' cancellation of the Fermi liquid renormalization of $\lambda^{-2}$ [24,25] 
does not occur in real superconductors which are not translationally (Gallalian) invariant. This is supported by experiment [26]. 
$n$ is proportional to the volume of the Fermi surface and according to Luttinger's theorem [41] 
it is independent of the strength of the correlations.  However, in iron-based superconductors where there are both electron and hole parts to the Fermi surface it should be noted that although the total volume is independent of the interactions, $n$ in the above formula may change. This is because although the total volume takes into account the sign of the carriers (it is zero for a compensated metal such as BaFe$_2$(As$_{1-x}$P$_x$)$_2$ where the electron and hole pockets have equal volume), the sign of $\bm{v}$ does not enter Eq.\:(\ref{Eq:lam}) and so the electron and hole pocket contributions add.  The dHvA results [10] 
suggest that although the volumes of the electron and hole pockets shrink as $x$ decreases from 1 towards the optimum composition, the main effect on $n/m^*$ near the QCP comes from the increase in $m^*$.   It should also be noted that $\lambda^{-2}$ may be reduced from the value derived from Eq.\:(\ref{Eq:lam}) by phase fluctuations [42], 
non-locality or disorder.



\end{document}